\documentclass[]{IEEEtran}

\usepackage{epsfig}

\usepackage{amsthm}
\usepackage[cmex10]{amsmath}
\usepackage{graphicx}
\usepackage{times}
\usepackage{latexsym}
\usepackage{graphicx}
\usepackage{amssymb}
\usepackage{amsfonts}
\usepackage{psfrag}
\usepackage{cite}

\newtheorem{thm}{Theorem}[section]

\newtheorem{lem}[thm]{Lemma}

\begin{document}

\title{Nested Polar Codes for  Wiretap and Relay Channels}

\author{Mattias Andersson, Vishwambhar Rathi, Ragnar Thobaben, J\"org Kliewer, and Mikael Skoglund}

\maketitle
\begin{abstract}
  We show that polar codes asymptotically achieve the whole
  capacity-equivocation region for the wiretap channel when the
  wiretapper's channel is degraded with respect to the main
  channel, and the weak secrecy notion is used. Our coding scheme
  also achieves the capacity of the physically degraded
  receiver-orthogonal relay channel. We show simulation results for
  moderate block length for the binary erasure wiretap channel,
  comparing polar codes and two edge type LDPC codes.
\end{abstract}

\maketitle
\vspace{-4mm}
\section{Introduction}
Polar codes were introduced by Arikan and were shown to
be capacity achieving for a large class of channels\cite{Ari09}.  Polar codes are
block codes of length $N=2^n$ with binary input alphabet $\mathcal X$.
Let $G = R F^{\otimes n}$, where $R$ is the bit-reversal mapping
defined in \cite{Ari09}, $ F =
  \begin{bmatrix}
    1 & 0 \\
    1 & 1
  \end{bmatrix},$ and $F^{\otimes n}$ denotes the $n^{\text{th}}$
  Kronecker power of $F$. Apply the linear transformation $G$ to $N$
  bits $\{u_i\}_{i=1}^N$ and send the result through $N$ independent copies of a binary
  input memoryless channel $W(y|x)$. This gives an $N$-dimensional
  channel $W_N(y_1^N|u_1^N)$, and Arikan's observation was that the
  channels seen by individual bits, defined by
\begin{align}
\label{eq:channel}
  W_N^{(i)}(y_1^N,u_1^{i-1}|u_i) = \sum_{u_{i+1}^N \in \mathcal{X}^{N-i}}\frac 1 {2^{N-1}} W_N(y_1^N|u_1^N),
\end{align}
\emph{polarize}, i.e as $N$ grows $W_N^{(i)}$ approaches either an
error-free channel or a completely noisy channel.

We define the polar code $P(N,\mathcal A)$ of length $N$ as follows. Given a subset $\mathcal A$ of the bits, set $u_i = 0$ for $i\in \mathcal A^{\mathcal C}$. We call $\mathcal A^{\mathcal C}$ the frozen set, and the bits $\{u_i\}_{i\in \mathcal A^{\mathcal C}}$ frozen bits. The codewords are given by $x^N = u_{\mathcal A}G_{\mathcal A}$, where $G_{\mathcal A}$ is the submatrix of $G$ formed by rows with indices in $\mathcal A$. The rate of $P(N,\mathcal A)$ is $|\mathcal A|/N$.

The block error probability using the successive cancellation (SC) decoding
rule defined by
\begin{align*}
  \hat{u}_i =
  \begin{cases}
    0 & i \in \mathcal A^{\mathcal C} \text{ or } \frac{W_N^{(i)}(y_1^N,\hat{u}_1^{i-1}|u_i = 0)}{W_N^{(i)}(y_1^N,\hat{u}_1^{i-1}|u_i = 1)} \geq 1 \text{ when }  i \in \mathcal A \\
    1 & \text{otherwise}
  \end{cases}
\end{align*}
can be upper bounded by $\sum_{i \in \mathcal A} Z_N^{(i)}$, where $Z_N^{(i)}$ is the Bhattacharyya parameter for the channel $W_N^{(i)}$\cite{Ari09}. It was shown in \cite{ArT09} that for any $\beta < 1/2$, 
\begin{align}
   \label{eq:polar}
   \liminf_{n \to \infty} \frac 1 N |\{i: Z_N^{(i)} < 2^{-N^\beta}\}| = I(W),
 \end{align}
where $I(W)$ is the symmetric capacity of $W$, which equals the Shannon capacity for symmetric channels. Thus if we let $\mathcal A_N = \{i: Z_N^{(i)} < 2^{-N^\beta}\}$, the rate of $P(N,\mathcal A_N)$ approaches $I(W)$ as $N$ grows. Also the block error probability $P_e$ using SC decoding is upper bounded by
\begin{align}
\label{eq:pe}
  P_e \leq N 2^{-N^\beta}.
\end{align}

We define the nested polar code $P(N,\mathcal A,\mathcal B)$ of length $N$ where $\mathcal B \subset \mathcal A$ as follows. The codewords of $P(N,\mathcal A,\mathcal B)$ are the same as the codewords for $P(N,\mathcal A)$. The nested structure is defined by partitioning $P(N,\mathcal A)$ as cosets of $P(N,\mathcal B)$. Thus codewords in $P(N,\mathcal A,\mathcal B)$ are given by $x^N = u_{\mathcal B} G_{\mathcal B} \oplus u_{\mathcal A \setminus \mathcal B} G_{\mathcal A \setminus \mathcal B}$, where $u_{\mathcal A \setminus \mathcal B}$ determines which coset the codeword lies in. Note that each coset will be a polar code with $\mathcal B^{\mathcal C}$ as the frozen set. The frozen bits $u_i$ are either 0 (if $i \in \mathcal A^{\mathcal C}$) or they equal the corresponding bits in $u_{\mathcal A \setminus \mathcal B}$.

Let $W$ and $\tilde W$ be two symmetric binary input memoryless
channels. Let $\tilde W$ be degraded with repect to $W$. Denote
the polarized channels as defined in (\ref{eq:channel}) by $W^{(i)}_N$  (resp. $\tilde W^{(i)}_N$), and their Bhattacharyya parameters by $Z^{(i)}_N$ (resp. $\tilde Z^{(i)}_N$). We will use the following Lemma which is Lemma 4.7 from \cite{Kor09}:

\begin{lem}
\label{lem:Korada}
If $\tilde W$ is degraded with respect to $W$ then $\tilde W^{(i)}_N$ is degraded with respect to $W^{(i)}_N$ and $\tilde Z^{(i)}_N \geq Z^{(i)}_N$.
\end{lem}
In Sections \ref{se:wiretap} and \ref{se:relay} we use Lemma \ref{lem:Korada} to show that nested polar codes are capacity achieving for the degraded wiretap channel and the physically degraded relay channel.

To our knowledge this work\footnote{This paper was originally submitted to this journal on March 5th, 2010.} is the first to consider polar codes for
the (degraded) relay channel. Independent recent work concerning the
wiretap channel includes \cite{MaVa10} and \cite{OzGa10}.

\vspace{-2.6mm}
\section{Nested Polar Wiretap Codes}
\label{se:wiretap}
We consider the wiretap channel introduced by Wyner\cite{Wyn75}. The sender, Alice, wants to transmit a message $S$
chosen uniformly at random from the set $\mathcal S$ to the intended
receiver, Bob, while trying to keep the message secure from a
wiretapper, Eve.  We assume that the input alphabet $\mathcal X$ is
binary, and Bob's output alphabets $\mathcal Y$ and Eve's output
alphabet $\mathcal Z$ are discrete. We assume that the main channel
(given by $P_{Y|X}$) and the wiretapper's channel (given by $P_{Z|X}$)
are symmetric. We also assume that $P_{Z|X}$ is stochastically
degraded with respect to $P_{Y|X}$, i.e.  there exists a probability
distribution $P_{Z|Y}$ such that $P_{Z|X}(z|x) = \sum_{y \in \mathcal
  Y} P_{Z|Y}(z|y)P_{Y|X}(y|x)$.

A codebook with block length $N$ for the wiretap channel is given by a
set of disjoint subcodes $\{\mathcal C(s) \subset \mathcal X^N\}_{s
  \in \mathcal S}$, where $\mathcal S$ is the set of possible
messages. To encode the message $s \in \mathcal S$, Alice chooses one
of the codewords in $\mathcal C(S)$ uniformly at random and transmits
it. Bob uses a decoder $\phi: \mathcal Y^N \to
\mathcal S$ to determine which message was sent.

A rate-equivocation pair $(R,R_e)$ is said to be achievable if $\forall \epsilon > 0$ and for a sufficiently large $N$, there exists a message set $\mathcal S$, subcodes  $\{C(s)\}_{s\in \mathcal{S}}$, and a decoder $\phi$ such that
\begin{align}
\label{eq:rate}
  \frac 1 N \log |\mathcal S| > R - \epsilon, \quad P(\phi(Y^N) \neq S) < \epsilon,
\end{align}
\begin{align}
   \frac 1 N \mathbb{H}(S|Z^N) > R_e - \epsilon, \label{eq:security}
\end{align}
where $ \mathbb{H}(S|Z^N)$ denotes the conditional entropy of $S$ given $Z^N$. The set of achievable pairs $(R,R_e)$ for this setting is 
\begin{equation}\label{eq:ach}
   R_e \leq R \leq C_M, \quad 0 \leq R_e \leq C_M - C_W,
\end{equation}
where $C_M$ is the capacity of the main channel, and $C_W$ is the capacity of the wiretapper's channel\cite{CsK78}.

In Theorem \ref{thm:1} we give a nested polar coding scheme\cite{LLPS07} for the wiretap channel that achieves the whole rate-equivocation rate region. Let the wiretapper's channel be denoted by $\tilde W$ and the main channel by $W$. We assume that $W$ and $\tilde W$ are symmetric, so $C_M = I(W)$ and $C_W = I(\tilde W)$.
\begin{thm}
\label{thm:1}
Let $(R,R_e)$ satisfy  (\ref{eq:ach}). For all $\epsilon > 0$ there exists a nested polar code of length $N=2^n$ that satisfies (\ref{eq:rate}) and (\ref{eq:security}) provided $n$ is large enough.
\end{thm}
\begin{IEEEproof}
Let $\beta < 1/2$, $\mathcal A_N = \{i: Z^{(i)}_N < 2^{-N^{\beta}}\}$, 
and let $\mathcal B_N$ be the subset of $\mathcal A_N$ of size $N(C_M - R)$ whose members have the smallest $\tilde Z^{(i)}_N$. Since (\ref{eq:polar}) implies $\liminf_{n \to \infty} |\mathcal A_N|/N = C_M \geq C_M - R$ such a subset exists if $n$ is large enough. This defines our nested polar code $P(N,\mathcal A_N,\mathcal B_N)$, and the subcodes $\mathcal C(s_N)$ are the cosets of $P(N,\mathcal B_N)$.

To send the message $s_N$, Alice generates the codeword
\begin{align}
\label{eq:subcode}
  X^N = T_N G_{\mathcal B_N} \oplus s_N G_{\mathcal A_N \setminus \mathcal B_N},
\end{align}
where $T_N$ is a binary vector of length $N(C_M-R)$ chosen uniformly at random. 

From (\ref{eq:pe}) the block error probability for Bob goes to zero as $n$ goes to infinity. The rate of the coding scheme is $\frac 1 N |\mathcal A_N \setminus \mathcal B_N|$, which goes to \mbox{$C_M - (C_M - R) = R$} as $n$ goes to infinity, since $\liminf_{n \to \infty} |\mathcal A_N|/N = C_M$. Thus our coding scheme satisfies (\ref{eq:rate}).

To show (\ref{eq:security}) we look at the equivocation for Eve. We first look at the case where $R\geq C_M - C_W$. We expand $I(X^N,S_N;Z^N)$ in two different ways and obtain
\begin{align}
  I(X^N,S_N;Z^N)  & = I(X^N;Z^N) + I(S_N;Z^N|X^N) \nonumber \\
 & = I(S_N;Z^N) + I(X^N;Z^N|S_N). \label{eq:michainrule}
\end{align}
Note that $I(S_N;Z^N|X^N) = 0$ as $S_N \to X^N \to Z^N$ is a Markov chain. By (\ref{eq:michainrule}) and noting 
$I(S_N;Z^N) =\mathbb{H}(S_N) - \mathbb{H}(S_N|Z^N)$, we write the equivocation rate $\mathbb{H}(S_N|Z^N)/N$ as  
 \begin{multline*}
\frac{\mathbb{H}(S_N) + I(X^N;Z^N|S_N) - I(X^N;Z^N)}{N} = \underbrace{\frac{\mathbb{H}(S_N)}{N}}_{=R-\delta(N)} + \\ 
\underbrace{\frac{\mathbb{H}(X^N|S_N)}{N}}_{=C_M - R} - \frac{\mathbb{H}(X^N|Z^N,S_N)}{N} - \underbrace{\frac{I(X^N;Z^N)}{N}}_{\leq C_W} \\
    \geq  C_M - C_W - \delta(N) - \frac{\mathbb{H}(X^N|Z^N,S_N)}{N},
 \end{multline*}
where $\delta(N)$ is the difference between $|\mathcal A_N \setminus \mathcal B_N|/N$ and $R$ which goes to zero as $n \to \infty$.

We now look at $\mathbb{H}(X^N|Z^N,S_N)$. For a fixed $S_N = s_N$ we see that $X^N \in \mathcal C(s_N)$. Let $P^{\prime}_e$ be the error probability of decoding this code using an SC decoder. By Lemma \ref{lem:Korada}, the set
$  \tilde{\mathcal{A}}_N = \{i: \tilde Z^{(i)}_N < 2^{-N^{\beta}}\} $ 
is a subset of $\mathcal A_N$. Also, $\liminf_{n \to \infty} \frac 1 N | \tilde{\mathcal{A}}_N| = C_W$, so if $|\mathcal B_N| \leq NC_W$ we have $\mathcal B_N \subset \tilde{\mathcal{A}}_N$ for large $n$, by the definition of $\mathcal B_N$. Since $|\mathcal B_N| = N(C_M - R) \leq NC_W$, we have $\tilde Z^{(i)}_N < 2^{-N^{\beta}} \ \forall i \in \mathcal B_N$ for large enough $n$. This implies 
$
P^{\prime}_e \leq \sum_{i \in \mathcal B_N} \tilde Z^{(i)}_N \leq N 2^{-N^{\beta}}.
$
We use Fano's inequality to show that $\mathbb{H}(X^N|Z^N,S_N)\to 0$:
\begin{align*}
  \liminf_{n \to \infty} \mathbb{H}(X^N|Z^N,S_N) &\leq \liminf_{n \to \infty} \left[\mathbb{H}(P^\prime_e) + P^\prime_e |\mathcal B_N|\right] = 0.
\end{align*}
Thus we have shown that
$
  \frac{\mathbb{H}(S_N|Z^N)}{N} \geq C_M - C_W - \epsilon \geq R_e - \epsilon
$
for $n$ large enough.

We now consider the case when $R < C_M - C_W$. The only difference from the analysis above is the term $\mathbb{H}(X^N|Z^N,S_N)$. Since $|\mathcal B_N| = N(C_M - R) > NC_W$, the code defined by  (\ref{eq:subcode}) is not decodable. Instead, let \mbox{$\mathcal B_{1N} = \{i:\tilde Z^{(i)}_N < 2^{-N^{\beta}}\}$}, $\mathcal B_{2N} = \mathcal B_N \setminus \mathcal B_{1N}$, and rewrite (\ref{eq:subcode}) as
$
  X^N =  T_{1N} G_{\mathcal B_{1N}} \oplus T_{2N}  G_{\mathcal B_{2N}} 
     \oplus S_N G_{\mathcal A_N \setminus \mathcal B_N}.
$

Note that, since $\liminf_{n \to \infty} |\mathcal B_{1N}|/N = C_W$, this code is decodable using SC given $T_{2N}$. If $T_{2N}$ is unknown we can try all possible combinations and come up with $2^{|\mathcal B_{2N}|}$ equally likely solutions (all solutions are equally likely since $T_N$ is chosen uniformly at random). Thus $\mathbb{H}(X^N|Z^N,S_N)$ should tend to $\mathbb{H}(T_{2N})$. We make this argument precise by bounding $\mathbb{H}(X^N|Z^N,S_N)$ as follows:
\begin{align*}
  \mathbb{H}(X^N|Z^N,S_N) &= \mathbb{H}(X^N,T_{2N}|Z^N,S_N) \\
  & = \mathbb{H}(T_{2N}|Z^N,S_N) + \mathbb{H}(X^N|Z^N,S_N,T_{2N}) \\
  &\leq \mathbb{H}(T_{2N}) + \mathbb{H}(X^N|Z^N,S_N,T_{2N})
\end{align*}
where in the last step we have used the fact that conditioning reduces entropy. We can show that the second term goes to zero using Fano's inequality as above. Since $\liminf_{n \to \infty}\frac{\mathbb{H}(T_{2N})}{N} = \liminf_{n \to \infty}\frac{|\mathcal B_{2N}|}{N} = C_M - R - C_W$, we get $\mathbb{H}(S_N|Z^N)/N \geq R - \epsilon$ for $n$ large enough.
\end{IEEEproof}
In Section \ref{se:relay} we show that the nested polar code scheme can be used to achieve capacity for the physically degraded receiver-orthogonal relay channel (PDRORC).
\vspace{-3mm}
\section{Nested Polar Relay Channel Codes}
\label{se:relay}
 The PDRORC is a three node channel with a sender, a relay, and a destination \cite{CoGa79}. The sender wishes to convey a message to the destination with the aid of the relay.
Let the input at the sender and the relay be denoted by $X$ and $X_1$ respectively, and let the corresponding alphabets $\mathcal X$ and $\mathcal X_1$ be binary. We denote the source to relay (SR) channel output by $Y_1$, the source to destination (SD) channel output by $Y'$, and the relay to destination (RD) channel output by $Y''$. We assume that the corresponding output alphabets $\mathcal Y_1, \mathcal Y'$, and $\mathcal Y''$ are discrete.
The SR and SD channel transition probabilities are given by $P_{Y'Y_1|X}$ and the RD channel transition probability is given by $P_{Y''|X_1}$. Note that the receiver components are orthogonal, i.e. $P_{Y'Y''|XX_1} = P_{Y'|X}P_{Y''|X_1}$. We further assume that the SD channel is physically degraded with respect to the SR channel, i.e $P_{Y'Y_1|X} = P_{Y_1|X}P_{Y'|Y_1}$, and that all the channels $P_{Y'|X}, P_{Y_1|X}$, and $P_{Y''|X_1}$ are symmetric. The capacity of the PDRORC channel is given by $C = \max_{p(x)p(x_1)}\min \left\{I(X;Y') + I(X_1;Y''),I(X;Y',Y_1)\right\}$. In the symmetric physically degraded case this simplifies to $C = \min \left\{C_{SD} + C_{RD},C_{SR}\right\}$, where $C_{SD}$, $C_{SR}$, and $C_{RD}$ are the capacities of the SD, SR, and RD channels respectively.
\begin{thm}
Let $R<C$. For all $\epsilon > 0$ there exists a nested polar code of rate $R$ and length $(B+1)N=(B+1)2^n$ such that the error probability at the destination is smaller than $\epsilon$ provided $B$ and $n$ are large enough.
\end{thm}
\begin{IEEEproof}
We use a block-Markov coding scheme and transmit $B$ codewords of length $N$ in $B+1$ blocks. Let $W$ and $\tilde W$ denote the SR and SD channels respectively. Let $Z^{(i)}_N$ and $\tilde Z^{(i)}_N$ be the Bhattacharyya parameters of the corresponding polarized channels.

First assume that $C_{SR} \leq C_{SD} + C_{RD}$. Let $\beta < 1/2$, $\mathcal A_N = \{i: Z^{(i)}_N < 2^{-N^{\beta}}\}$, and let $\mathcal B_N= \{i: \tilde Z^{(i)}_N < 2^{-N^{\beta}}\}$. By Lemma \ref{lem:Korada}, $\mathcal B_N \subset \mathcal A_N$. The source will transmit in each block using the nested polar code $P(N,\mathcal A_N,\mathcal B_N)$. After receiving the whole codeword the relay decodes the bits in $\mathcal A_N$. The probability that the relay makes an error when decoding can be made smaller than $\epsilon/(3B)$ by choosing $n$ large enough. The relay then reencodes the bits in $\mathcal A_N \setminus \mathcal B_N$ and transmits them using a polar code of rate $(|\mathcal A_N| - |\mathcal B_N|)/N$ in the next block. In general, in block $k$ the source transmits the $k^{\text{th}}$ codeword while the relay transmits the bits in $\mathcal A_N \setminus \mathcal B_N$ from the $(k-1)^{\text{th}}$ block. The destination first decodes the bits in $\mathcal A_N \setminus \mathcal B_N$ using the transmission from the relay. This can be done with error probability smaller than $\epsilon/(3B)$ provided $n$ is large enough since the rate of the relay to destination code tends to $C_{SR}-C_{SD} \leq C_{RD}$ as $n$ grows.
Finally the destination decodes the source transmission from the $(k-1)^{\text{th}}$ block. It uses the bits from the relay transmission in block $k$ to determine which coset of $P(N,\mathcal B_N)$ the codeword lies in. The rate of $P(N,\mathcal B_N)$ approaches $C_{SD}$ so the destination can decode with block error probability smaller than $\epsilon/(3B)$. By the union bound the overall error probability over all $B$ blocks is then smaller than $\epsilon$. The rate of the scheme is $B|\mathcal{A}_N|/N(B+1)$ which can be made arbitrarily close to $C_{SR}$ provided $B$ and $n$ are large enough since $\liminf_{n\to \infty} |\mathcal A_N|/N = C_{SR}$.

Now assume that $C_{SR} > C_{SD} + C_{RD}$. Let $\mathcal B_N = \{i: \tilde Z^{(i)}_N < 2^{-N^{\beta}}\}$ and let $\mathcal A_N$ be a subset of $\{i:Z^{(i)}_N < 2^{-N^\beta}\}$ of size $N(C_{SD} + C_{RD})$ containing $\mathcal B_N$. Such a subset exists provided $n$ is large enough since $C_{SR} > C_{SD} + C_{RD}$. The analysis of the block error probability is the same as in the first case, and the rate of the coding scheme is $B|\mathcal{A}_N|/N(B+1)$ which approaches $C_{SD} + C_{RD}$ when $n$ and $B$ are large.
\end{IEEEproof}
\vspace{-3mm}
\section{Simulations}
\label{sec:sim}
\begin{figure}[!t]
  \centering
  \includegraphics[width=\columnwidth]{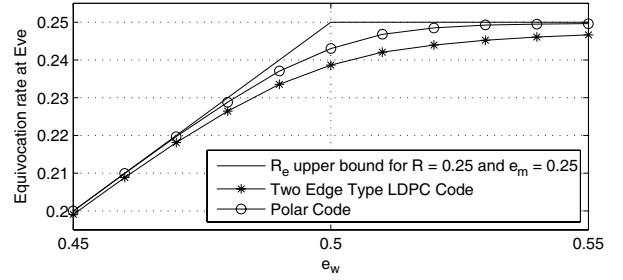}
  \vspace{-6mm}
\caption{Equivocation rate versus $e_w$. Codes designed for \mbox{$R = 0.25$}, \mbox{$e_m = 0.25$}, \mbox{$\ e_w = 0.5$}, and block length $N = 1024$.}
\vspace{-5mm}
\label{fig:eq}
\end{figure}

We show simulation results comparing Eve's equivocation for nested polar wiretap codes and two edge type LDPC codes over a wiretap channel where both the main channel and the wiretapper's channel are binary erasure channels with erasure probabilities $e_m$ and $e_w$ respectively. The LDPC codes are optimized using the methods in \cite{RATKS09} and for the LDPC codes the curve shows the ensemble average. The equivocation at Eve is calculated using an extension of a result in\cite{OzW84}\footnote{Note that the polar codes $P(N,\mathcal A_N)$ and $P(N,\mathcal B_N)$ are linear codes and we therefore can calculate the corresponding parity check matrices.}:
\begin{lem}
Let $H$ be a parity check matrix for the overall code ($P(N,\mathcal A_N)$ in the polar case) and let $H^{(s)}$ be a parity check matrix for the subcode $(P(N,\mathcal B_N))$ in a nested coding scheme for the binary erasure channel. Then the equivocation at Eve is
$   \text{rank}(H_{\mathcal E}^{(s)}) - \text{rank}(H_{\mathcal E})$, where $H_{\mathcal E}$ is the matrix formed from the columns of $H$ corresponding to erased codeword positions.
\end{lem}
\begin{IEEEproof}
  The equivocation at Eve can be written as
\begin{IEEEeqnarray}{rCl}
  \mathbb{H}(S_N|Z^N) &= &\mathbb{H}(X^N|Z^N) - \mathbb{H}(X^N|S_N,Z^N).
\end{IEEEeqnarray}
For a specific received $z$ we have
$
  H_{\mathcal E}x_{\mathcal E}^T + H_{\mathcal E^{\mathcal C}} x_{\mathcal E^{\mathcal C}}^T = 0,
$
where $x_{\mathcal E}^T$ is unknown. The above equation has $2^{N-\text{rank}(H_{\mathcal E})}$ solutions, all of which are equally likely since the original codewords $X^N$ are equally likely. In the same way $\mathbb{H}(X^N|S_N,Z^N) = N-\text{rank}(H_{\mathcal E}^{(s)})$. This implies 
$
  \mathbb{H}(S_N|Z^N) = \text{rank}(H_{\mathcal E}^{(s)}) - \text{rank}(H_{\mathcal E}).
$
\end{IEEEproof}
Fig. \ref{fig:eq} shows the equivocation rate at Eve, and also the upper bound for $R_e$ as a function of $e_w$ for fixed $R=0.25$ and $e_m=0.25$. It is interesting to note that even with a block length of only 1024 bits the curves are close to the upper bound.
\vspace{-4mm}
\section{Acknowledgement}
We wish to thank an anonymous reviewer for pointing out the existence
of the related preprints \cite{MaVa10} and \cite{OzGa10}.
\bibliographystyle{IEEEtran} 
\vspace{-3mm}
\bibliography{IEEEabrv,Andersson}

\end{document}